\documentclass[%
reprint,
aps,
floatfix,
]{revtex4-1}

\usepackage{ulem}

\usepackage{amsmath}    
\usepackage{graphicx}   
\usepackage{verbatim}   
\usepackage{color}      
\usepackage{subfigure}  
\usepackage{hyperref}   
\graphicspath{{./}{figs/}}

\begin{document}

\begin{abstract}
We observe matterwave interference of a single cesium atom in free fall.  The interferometer is an absolute sensor of acceleration and we show that this technique is sensitive to forces at the level of $3.2\times10^{-27}$~N with a spatial resolution at the micron scale.  We observe the build up of the interference pattern one atom at a time in an interferometer where the mean path separation extends far beyond the coherence length of the atom.  Using the coherence length of the atom wavepacket as a metric, we directly probe the velocity distribution and measure the temperature of a single atom in free fall.
\end{abstract}

\title{Observation of Free-Space Single-Atom Matterwave Interference}
\author{L. P. Parazzoli}
\affiliation{Sandia National Laboratories, Albuquerque, New Mexico 87185, USA}
\author{A. M. Hankin}
\author{G. W. Biedermann}
\affiliation{Sandia National Laboratories, Albuquerque, New Mexico 87185, USA}
\affiliation{Center for Quantum Information and Control (CQuIC), Department of Physics and Astronomy, University of New Mexico, Albuquerque, New Mexico 87131, USA }
\date{\today}

\maketitle

Over the past 20 years light-pulse atom interferometers have shown an exceptional capacity for precision metrology.  Demonstrations have been performed in a wide variety of domains from practical applications in inertial sensing~\cite{mcguinness2012, Canuel06, pyramid} to advancing foundational knowledge in physics~\cite{muller_nature_2010, gravwave, Fixler, Dimopoulos07}.  To maximize sensitivity, the majority of atom interferometers utilize large ensembles of atoms or high flux beams.  Even so, it is generally accepted that atom interferometers can operate because the atoms interfere with themselves~\cite{Pritchard2009}, as has been demonstrated with electrons~\cite{kawasaki1989, Gossard_singleElectron} and neutrons~\cite{bonse1974}.  Despite its patency, this self-interference phenomenon has never been directly observed, especially in a system where the path separation extends far beyond the coherence length of the particle~\cite{toschek_AI_singleAtom}.  This is due in large part to the experimental challenges associated with single neutral atom trapping, control, and detection.  Inspired by recent advances~\cite{Grangier_subPoissanLoading}, we use a micron-scale optical tweezer to observe a single cesium atom in a light-pulse atom interferometer experiment where the wavepacket separation is 240 times larger than the coherence length.  In doing so, we also introduce a technique to probe forces with high spatial resolution that inherits the absolute accuracy intrinsic to atom interferometry.

For most applications of atom interferometry, a {\it bulk}-atom interferometer approach is well suited.  However, a significant advantage of implementing an atom interferometer using a single atom in an optical tweezer is that the atom itself can be highly localized in space.  Of particular interest at this length scale is the ability to probe, with absolute accuracy, forces that are very near to surfaces~\cite{gerci2003prd} such as Casimir-Polder forces~\cite{casimir_polder_1948, hinds_1993} as well as hypothetical forces that result in  non-relativistic deviations from Newtonian gravitation~\cite{Dimopoulos1996, dimopoulos1998.b,Dimopoulos1998.a}.  Predicted to appear at sub-millimeter length scales, direct observation of non-Newtonian gravity could lead to the validation of physics beyond the Standard Model and to a unification of gravity and quantum theory.  The measurement of Casimir-Polder forces has important applications to stiction in micro components and  verifying complex theoretical calculations for non-trivial surface geometries.

\begin{figure}
\includegraphics[width=3.5 in]{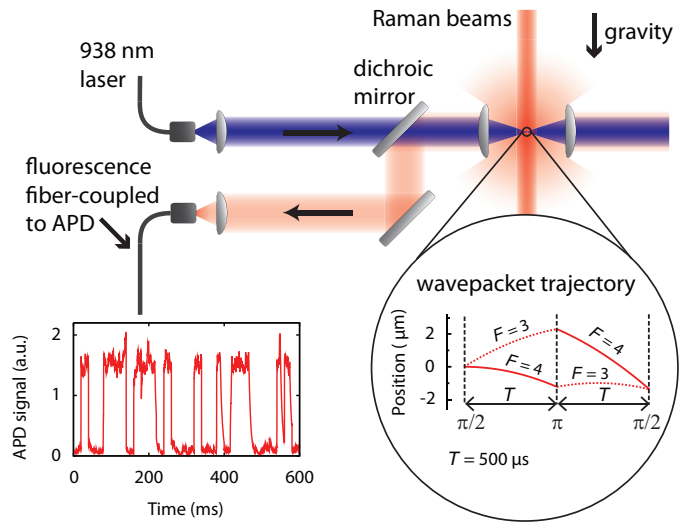}
\caption{\label{fig1} Diagram depicting the apparatus for observing a single-atom interferometer.  A single atom is trapped in an optical tweezer.  The florescence from the atom is coupled to an avalanche photodiode (APD) for detection, showing (bottom left) the two discrete levels of photon counts that are characteristic of collisionally blockaded loading of single atoms into an optical tweezer.  A wavepacket trajectory is shown for an atom in free-fall under the influence of gravity and a light pulse atom interferometer sequence. }
\end{figure}

\begin{figure}[]
\includegraphics[width=3.75 in]{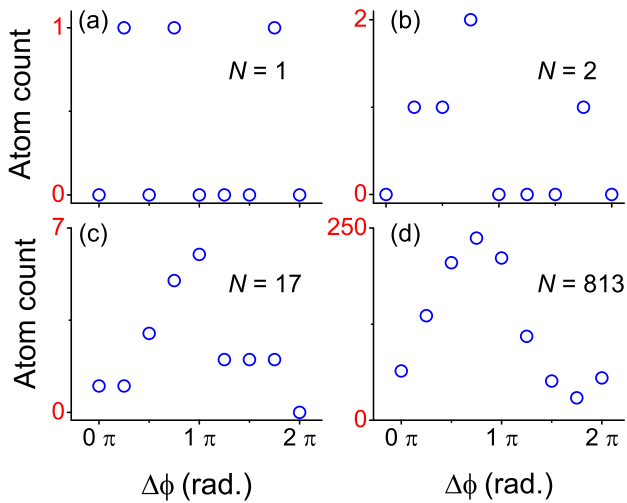}
\caption{\label{fig2} Emergence of the interference fringe for $T=74.5~\mu$s.  The plots show the cumulative number of Cs atoms (per phase) detected in $\vert F=3\rangle$ after $N$ independent single atom experiments for (a) $N$=1, (b) $N$=2, (c) $N$=17, and (d) $N$=813.  Note individual scaling of vertical axes. }
\end{figure}

In this Letter, we demonstrate a single-atom interference signal by creating a Mach-Zehnder type interferometer with a cesium atom that has been isolated by an optical tweezer and released into free-fall.  The analogue of the ``beam splitters'' and ``mirrors'' in the interferometer are created by light pulses that drive stimulated Raman transitions as illustrated in the inset of Fig.~\ref{fig1}.  The stimulated Raman transition entangles the two hyperfine states of the atom with two momentum states that are separated by $\hbar k_\textrm{eff}$, where $k_\textrm{eff}$ is the wavevector of the Raman field.  The atom then evolves in a coherent superposition of position states separated by $\Delta x(t) = \hbar k_\textrm{eff}t/m$, where $m$ is the mass of the cesium atom and $t$ is the time of the evolution.  In our experiment the position states separate by as much as 3.5~$\mu$m after the first Raman pulse.  Subsequent pulses redirect the atomic wavepackets back toward each other and then recombine them.  After the wavepackets are recombined, the atom is recaptured in the optical tweezer to measure the differential phase shift of the two paths.  Figure~\ref{fig2} details the build-up process of the interference pattern one  atom at a time.  The emergence of this interference pattern relies on the precise recombination of the wavepackets at a level given by the coherence length, which we measure to be 14.8~nm.

The apparatus for trapping the single atom consists of an optical dipole trapping laser~\cite{tweezerLoad} focused through a reservoir of cold atoms produced by a magneto-optical trap (MOT).  This takes place within an ultra-high vacuum apparatus enabling a background-limited trap lifetime of 10-15~s.  The optical dipole trap is tuned to 938 nm near the magic wavelength for the $6S_{1/2} \rightarrow 6P_{3/2}$ resonance of atomic cesium~\cite{Cs_magicWL}, coinciding with the laser cooling transition.  Trapping at the magic wavelength allows the atoms from the MOT to be readily cooled into the optical tweezer. The trapping light focuses to a 1/$e^2$ radius of 1.8 $\mu$m through a 2.75~mm focal length molded glass aspheric lens.  The lens is mounted inside the vacuum apparatus to avoid aberrations that arise from focusing through a glass plate.  The trap size is made to be sufficiently small such that we operate in the collisional blockade regime~\cite{grangier2002blockade}, which ensures that no more than one atom can be loaded into the trap. The same lens which is used to create the trap is also ideally suited to collect fluorescence from the atom within the trap with very high spatial-discrimination.  Once an atom enters the trap, light-induced-fluorescence from the cooling transition at 852~nm is coupled backward through the trapping lens and separated from the trapping laser by a dichroic mirror.  To minimize background light during detection, the fluorescence is coupled into a single mode fiber before it is detected by an avalanche photodiode (APD).  Using this method, we obtain a clearly resolved single-atom fluorescence signal relative to background.  This fluorescence signal is monitored in real-time and a threshold fluorescence level is set to indicate the successful loading of the trap and to trigger the start of an experimental cycle.

After loading the trap, we further cool the trapped atom in a two stage sequence.  Initially, the trapped atom is cooled from $36~\mu$K to $10~\mu$K using sub-Doppler cooling.  During the sub-Doppler cooling, we linearly ramp the MOT laser detuning from 3.3~$\Gamma$ to 10.9~$\Gamma$ (where $\Gamma = 2\pi\cdot5.234$~MHz is the natural linewidth of the $6S_{1/2} \rightarrow 6P_{3/2}$ transition), while simultaneously ramping down the intensity from $I\approx 3.6~\textrm{mW/cm}^2$ (per beam) to approximately half the initial value.  The duration of the sub-Doppler cooling sequence is 1.2~ms.  We then further reduce the temperature of the atom to $4.2~\mu$K with an adiabatic cooling process~\cite{grangier_cooling_2011}.  This is achieved by switching off the MOT light and then adiabatically ramping down the trapping laser peak intensity from 64~kW/cm$^2$ $(U_\textrm{trap} = 550~\mu\textrm{K})$ to a final intensity of 4.7~kW/cm$^2$ $(U_\textrm{trap} = 40~\mu\textrm{K})$ over a 2~ms period.  The root-mean-square (rms) velocity of the cesium atom at this temperature is $v_\textrm{rms}=$16.2~mm/s.

After cooling, the interferometer is initialized by optically pumping the atom to the upper ``clock state'', $\vert F=4, m_F=0 \rangle$.  The optical pumping is implemented using $\pi$-polarized light in a $\sim3$ Gauss magnetic field and tuned to $\vert F=4 \rangle \rightarrow \vert F'=4 \rangle$ of the D1 manifold.  The atom is simultaneously illuminated with non-polarized repump light tuned to $\vert F=3\rangle \rightarrow \vert F'=4 \rangle$ of the D2 manifold.  Through this method, we achieve a total population transfer into  $\vert F=4, m_F=0\rangle$ of 96\%.

Following state preparation we extinguish the optical dipole trap, releasing the atom into free-fall, and apply the aforementioned interferometer sequence.  The Raman fields are pulsed in a $\pi/2-\pi-\pi/2$ sequence with each pulse separated by the interrogation time $T$. We then recapture the atom in the same trap with a probability that depends on the temperature, trap size, and time-of-flight.  To maximize the recapture probability, we recapture using full trapping laser intensity (i.e. same trapping laser intensity as prior to adiabatic cooling).  We then project the matter-wave superposition using a blast pulse to remove the atom from the trap if it projects into $\vert F=4 \rangle$ and then detect the atom if it projects into $\vert F=3 \rangle$. The atom is detected by flashing on the MOT laser for 5~ms and collecting the fluorescence as described before.  The measurement cycle is then repeated by again collecting a reservoir of cold atoms in an overlapping MOT volume and loading a new atom.  In this way we determine the interferometer phase shift by averaging single atom experiments under identical conditions.

\begin{figure}
\includegraphics[width=3.75 in]{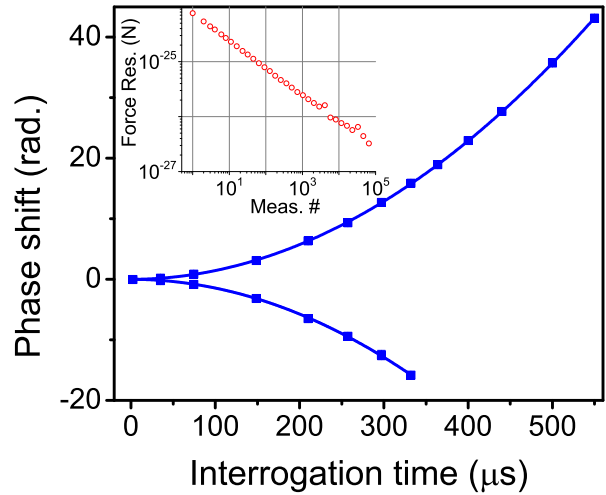}
\caption{\label{fig3} Atom interferometer phase evolution as a function of interrogation time for two cases: $k_\textrm{eff}$ pointed along (upper curve) and opposite (lower curve) the direction of gravity. The measured phase shift (blue squares) is determined at each interrogation time by measuring the interferometric fringe, as in Fig.~\ref{fig2}.  The points are fit (blue line) to determine the acceleration of the atom due to gravity.  The inset shows a two-sample Allan deviation of the force resolution as a function of the number of measurements for $T=363.9~\mu$s.}
\end{figure}

Owing to the homogeneity of the gravity field, the response of the interferometer is determined primarily, not by the path integral, but by the interaction of the atom with the Raman field such that the probability to measure the atom in the state $|F=3, m_F=0\rangle$ is given by $P_{|F=3\rangle} = \frac{1}{2}(1-\cos\Delta\phi)$, where~\cite{peters}
\begin{eqnarray}
\Delta\phi = k_\textrm{eff}g_k\left[ T^2 + t_\pi\left(1+\frac{2}{\pi}\right)T \right]\label{eq.ai}\label{eq1}.
\end{eqnarray}
Here $t_\pi = 1.0~\mu\textrm{s}$ is the length of the $\pi$-pulse and $g_k = g\cos(\theta)$ is the projection of $k_\textrm{eff}$ onto gravity.  The term that is linear in $T$ is a small corrective term to compensate for the Doppler shift of the Raman lasers as the atom accelerates with gravity.  The phase offset $\Delta \phi$ is measured by scanning the phase of the Raman coupling field after the first $\pi/2$ pulse to reveal the interferometer fringe (Fig.~\ref{fig2}).  This measurement is repeated for several values of the interrogation time and the resulting phase evolution is shown (blue squares) in Fig.~\ref{fig3}. The points are fit (solid line) to the predicted phase evolution (Eq.~\ref{eq1}), yielding a value of $g=9.8$~m/s$^2$.  The measurement is made for $\theta = 10^\circ$ (upper curve) and $\theta = 190^\circ$ (lower curve) relative to the direction of gravity.  As expected, reversing the direction of $k_\textrm{eff}$ also reverses the sign of the phase shift.

The sensitivity of the interferometric measurement scales quadratically with the interrogation time.  However, because the atom is not confined during the measurement, as the interrogation time increases, the probability that the atom is recaptured simultaneously decreases, resulting in a loss of signal.  In our experiment, the interrogation time is limited by the temperature~\footnote{ For clarification, while the {\it temperature} of a single atom is not a well defined measurement, in this context we define it as a measure of the average energy (or momentum spread) of single atoms realized over many iterations.} of the atom and the trap size. Although an optimal interrogation time may be chosen based on the recapture probability, the ultimate sensitivity is obtained by minimizing the temperature of the atom.  Temperature measurements in an optical tweezer relying on ballistic expansion~\cite{grangier_cooling_2011} can be obfuscated by uncertainty in the trap profile.  Accordingly, we demonstrate a technique to measure the temperature of a single atom in free space.  This method advances an interferometric approach to probe the mean longitudinal coherence length of the atom which can be related to the momentum dispersion via Heisenberg's uncertainty principle.  This is the first demonstration of a {\it direct} free-space measurement of single neutral atom temperature, independent of external factors such as trap geometry.

\begin{figure}
\includegraphics[width=3.5 in]{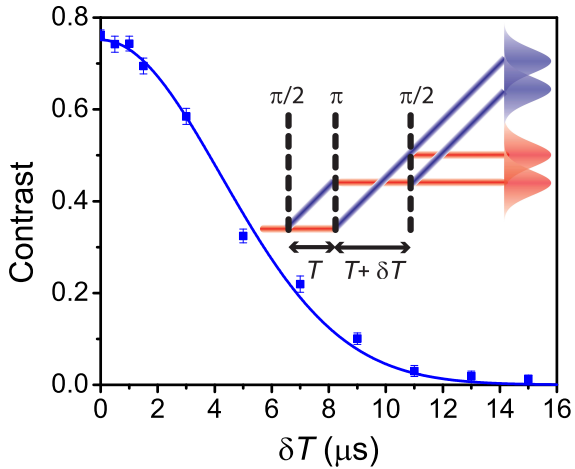}
\caption{\label{fig4} Measuring atom temperature by probing coherence length.  The inset shows a $\pi/2-\pi-\pi/2$ sequence whose temporal symmetry is broken by the addition of a delay time $\delta T$ before the final pulse.  The contrast of the interference fringe (squares) is measured for each $\delta T$ and fit (solid line) to the convolution of two Gaussians to give the wavepacket coherence length.}
\end{figure}

The interferometer sequence shown in Fig.~\ref{fig4} is used to measure the atom's coherence length in free space.  This sequence is nearly identical to the one described previously, using a Raman coupling field to split, reflect, and recombine the atomic wavepackets.  The difference in this case is that the second interrogation time is extended by a duration of $\delta T$, such that the wavepackets do not recombine perfectly, having now a separation of $\Delta x = v_r\delta T$, where $v_r$ is the two-photon recoil velocity.  As the separation is stepped to larger values the fringe contrast decreases monotonically with a characteristic distance set by the coherence length of the atom.  It has been demonstrated that the coherence length remains constant as the free-space wavepackets evolve in time~\cite{Kellog_longCoherence}.  The fringe contrast, $\chi(\delta T)$, is then given by the convolution of the free-space Gaussian wavepackets defined by the coherence length $x_a$,
%
\begin{eqnarray}
\chi(\delta T) = \exp\left( -\frac{v_r\delta T^2}{8 x_a^2} \right)\label{eqn:contrast}.
\end{eqnarray}
%
Figure~\ref{fig4} (blue squares) shows how the contrast decays with $\delta T$.  Fitting to the above equation (blue line) gives a coherence length of 14.8(4)~nm that, from $\Delta x \Delta p = \hbar/2$, corresponds to a velocity uncertainty of 16.2(5)~mm/s, and a temperature of 4.2(2)~$\mu$K.  We also measure the in-trap temperature of the atom to verify that there is no heating of the atom while the trap switches off.  The in-trap measurement is made using Doppler-sensitive Raman spectroscopy with a Doppler-limited Rabi rate of $\Omega = 12.5$~kHz.  After accounting for the Raman pulse width, we find a Gaussian transition width of 38.5(5)~kHz corresponding to a temperature of 4.3(1)~$\mu$K, showing good agreement with the free-space temperature measurement.

This measured temperature is an order of magnitude greater than the limit imposed by the ground state energy of the trap, encouraging further development~\cite{wineland_1989}. Nevertheless, we demonstrate a force sensitivity useful for measurements of Casimir-Polder potentials.  The inset of Fig.~\ref{fig3}  shows the Allan deviation for $T=363.9~\mu$s, where the competing effects of increasing $T^2$ sensitivity and decreasing recapture probability give optimal signal-to-noise.  The force sensitivity of the interferometer after $10^5$ measurements at this interrogation time is $3.2\times10^{-27}$~N.  With this sensitivity Casimir-Polder forces become measurable at distances below 3~$\mu$m.  We note that, after accounting for the fraction of atoms lost during free-flight, our observed sensitivity agrees with that predicted by quantum projection noise.  The technique is stable against long term drift as witnessed by the $N^{-1/2}$ trend in the Allan deviation for $10^5$ measurements.  This data set represents 54 hours of continuous measurements with a mean repetition rate of $1.5$~Hz.  Enhancing the sensitivity by two orders of magnitude would enable further constraints on non-Newtonian gravity at the micron length scale.  Such enhancements can follow from a combination of ground state cooling~\cite{sideBand} and an increase in the experimental data rate through lossless detection schemes~\cite{browaeys_detection2011, chapman2011}.

In conclusion, we have observed matterwave interference of a free-space single-atom that is initially confined in an optical tweezer.  In contrast to conventional atom interferometer approaches, which utilize large ensembles of trapped atoms, an interferometer consisting of a single atom allows for the measurement of highly localized forces such as those existing very close to material surfaces.  Measuring localized forces near surfaces has important applications in the characterization of materials and in studying the fundamental laws of physics.  We also demonstrate an alternate method to determine the temperature of a single atom trapped in an optical tweezer by measuring the free-space coherence length.  This technique provides an accurate measurement of the atom temperature independent of the trapping potential.

We would like to thank Y. Jau, M. Mangan, C. Johnson, P. Schwindt, G. Burns, and A. Geraci for discussions and support in the development of the experiment.  This work was supported by the Laboratory Directed Research and Development program at Sandia
National Laboratories.

%

\end{document}